\documentstyle{article}
\input psfig
\long\def\comment#1{}
\begin{document}
\title{Quantum Key Distribution Using Three Basis States}
\author{Subhash Kak\\
Department of Electrical \& Computer Engineering\\
Louisiana State University\\
Baton Rouge, LA 70803-5901; {\tt kak@ee.lsu.edu}}
\maketitle

\begin{abstract}
This note presents a method of public key distribution using
quantum communication of $n$ photons that simultaneously provides a high
probability that the bits have not been tampered.
It is a variant of the quantum method of Bennett and Brassard (BB84)
where the transmission states have been decreased from 4 to 3
and the detector states have been increased from 2 to 3.
Under certain assumptions regarding method of attack,
it provides superior performance (in terms of the number of usable
key bits) for $n < 18 m$, where $m$ is the number of key bits
used to verify the integrity of the process in the BB84-protocol.



\end{abstract}

\section{Introduction}

Since the act of measurement reduces the state function
of a quantum object,
one cannot determine what state characterized
the system prior to the measurement.
This property is at the heart of the method of
quantum key distribution which,
under certain assumptions, provides an absolutely secure method
of sharing a secret subject to public exchange of side information.
As an example of quantum communication, the
problem of key distribution
forms a part of the general subject of
quantum information science that has relevance
for our understanding of the foundations of
quantum and information theories.

In cryptography, two users, by convention called
Alice (A) and Bob (B), wish to exchange
information which must remain inaccessible
to Eve (the eavesdropper).
The quantum key distribution method 
of Bennett and Brassard (1984) (BB84), presented first
at a conference in
Bangalore sixteen years ago, uses
4 different polarizations of the photons and a pair
of basis states in the detector for this
information exchange.
The BB84 protocol is symmetric in its use of 
the polarizations.
After the key has been obtained, this protocol requires the exchange
of further information about parity of randomly chosen subsets of the
key.

Can we devise a scheme, somewhat in the spirit of joint encryption
and error-correction coding (Kak 1985), that will not only distribute
the key but also provide additional information about the
integrity of the distribution process?
In particular,
we would like to have a method where the additional exchange
of information, that is required after
the key has been distributed, is unnecessary.

Given the symmetry of the BB84 method, one would expect that
it could not be improved upon.
But symmetry-breaking can provide advantage for part
of the range of the operation of the system,
albeit at the cost of some leakage of information. 
In this note we consider a variant protocol that
breaks the symmetry
of the BB84 method.
In our method, three polarization states
of the photons and three basis states of the detector are used.
We show that doing so provides advantage over
the BB84 method for certain sizes of the key under certain, not unreasonable,
assumptions about the attacks mounted on it.

\section{The BB84 protocol}

In the BB84 method
Alice (A) chooses photons (or other particles)
prepared in 
four polarization states 
of $0,~45,~90$ and $135$ degrees and sends $n$ of them in random order to Bob (B),
who measures each photon using detectors matched in a random
sequence to two pairs of
orthogonal bases: $(0,~90)$ and $(45,~135)$ degrees.
Note that if the directions and the polarizations of
$0,~45,~90$ and $135$ degrees are represented by $0,~k,~1$ and $l$, respectively,
it is sufficient to have the directions $0$ and $k$ as
the bases in the detector.
If it is assumed that the photons are sent according to a clock, when
the detector outputs nothing ($e$), it is clear that the input was a
photon in a state orthogonal to the detector setting.

Bob now tells Alice the sequence of the bases he used for his
detection. Alice informs him which detector bases were correctly
aligned. Alice and Bob keep only the data from these correctly
measured (or inferred) photons, discarding the rest.
Now Alice and Bob test their key by publicly choosing a random
subset of bit positions and verifying that this subset
has the same parity (defined as odd) in their respective
versions of the key. If their keys had
differed in one or more bit positions, this test
would have discovered that fact with probability of
$\frac{1}{2}$. One bit is now discarded, to compensate
for the information leaked when its parity was revealed. This step
is repeated $m$ times, leading to a certification with
probability of $1~-~2^{-m}$ that their mutual keys are identical,
at the cost of reducing the key by $m$ bits.
Since the average number of correct alignments between 
the input bases and the detection pairs is
about half of the total number of photons sent,
the expected size of the key, certified with
the probability 
$1-2^{-m},$ is $\frac{n}{2} - m$.

This protocol assumes that Eve, the eavesdropper, cannot
make multiple copies of the transmitted photons, perform
various tests on them, and estimate
their original polarizations using the public channel information
from Alice. 
Should Eve intercept the photons directly, make
measurements and transmit the measured
photons to Bob, it would change the polarizations
and this would be detected by Alice and Bob
in their exchanges of the parity of the random subsets of the
exchanged key.

It is interesting that there is a certain complementarity
in the workings of classical and quantum communication.
The state of a classical object can be fully determined and
duplicated leading to ease of storage and
transmission, and difficulty of secrecy-coding. On the other hand, the 
state of the quantum object is not fully known, so
it is hard to duplicate it and hard to store and transmit it,
while secrecy-coding is easy.
The information associated with classical
and quantum processing is also different (Kak 1998).

\section{Key distribution with three states}

In our asymmetric method,
photons are prepared by Alice in the polarization states of
of $0,~45$ and $90$ degrees only.
At the receiving end, Bob uses filters before his detector that are
matched to the same polarizations.
We have introduced asymmetry at two places: by cutting down
on the number of photon polarizations from 4 to 3, and by
increasing the number of detector states from 2 (which is
equivalent to 4) to 3.

Table 1 summarizes the 9 different possibilities between the
photon and the detector states for the data from Alice,
Bob's detector settings, and what Bob actually receives.

\vspace{2.5mm}
{\em Table 1}: Photon and detector states
\vspace{2mm}

\begin{tabular}{||l|ccccccccc||}  \hline
Alice's data & 0 & 0 & 0 & k & k & k & 1 & 1 & 1 \\ \hline
Bob's filter   & 0 & k & 1 & 0 & k & 1 & 0 & k & 1 \\ \hline
Bob receives & 0 & k/e & e & 0/e & k & 1/e & e & k/e & 1 \\ \hline
\end{tabular}

\vspace*{3.5mm}

Bob sends information about
his filter settings to Alice on a public channel,
and Alice uses the same public channel to tell Bob which 
settings were correct.
The latter information makes it possible for Bob 
to infer which $e$ outputs should
be replaced by either a $0$ or $1$.
For example, if Bob is told by Alice that his setting
``1'' is correct in the 3rd column of the Table, he would
know that he should replace his estimate of 
the transmitted polarization from ``nothing''  ($e$)
to $0$.

As we can see, out of these 9 cases Bob is able to 
correctly receive or infer 5 cases. 
Now 
Eve, the eavesdropper, 
being unable to intercept or duplicate the photons being
sent by
Alice to Bob, can only 
work on the information being exchanged on
the public channel.
The only case where she would know the correct
output is when Bob's filter setting
is k and it is declared correct by Alice.
This means that she knows one of the 5 correct
cases of Bob.
So while the correct recognition rate for Bob is $\frac{5}{9}$,
bits at a rate of only $\frac{4}{9}$ have guaranteed security.
Just these 4 bits out of the 9 listed in the
Table will actually be used for creating the
key, the fifth bit (the setting of k) helps in authenticating the integrity
of the process. The latter bits, at
the rate of $\frac{1}{9}$, provide
confirmation that there has been no tampering with
the transmitted photons. 

There is a 
$1/3$ probability that Eve would have used the
correct filter placement at that spot if she had intercepted
the photon sequence and replaced it with her own, so the
transmission for every 9 bits can be certified to the probability
$1-1/3 = 2/3$.

Quantum information is properly examined only in terms of
the arrangement of the system and the experimenter.
We assume here that
Eve, like Bob, is using a receiver with settings of 3
basis states with the specified directional
information. The
certification probabilities quoted in this paper are based on this
assumption.
Unlike the BB84-method, Eve can, by using filter settings
stuck at $0$ or $1$, know the polarization of one-third of the 
transmitted photons. But that way, she could never hope
to guess all the polarization choices made by Alice,
and so this would not constitute a practical method of attack.
To understand this clearly, assume that Eve's filter settings
are all at 0. She would receive a transmitted 0 as 0, a
transmitted 1 as e, and a transmitted k as 0 or e. One half
of her received bits are 0 and the other half are 1, and 
she is unable to work backwards and guess any specific
value from her data alone.

\subsection*{A probabilistic analysis}
Prior
to being told the correct locations, the average mutual information,
$I(A;B)$, between Alice and Bob 
is computed by
\[I(A;B) = H(A) + H(B) - H(A, B).\]
where 
$H(A) , H(B)$, and $ H(A, B)$ are the individual entropy
measures of $A$, $B$, and the joint entropy of $A$ and $B$,
respectively.

$H(A)$ equals $log 3 \approx 1.585$ bits. $H(B)$ is computed by first
finding the probabilities of the four received states,
$0, k, 1, e$, which are  easily seen to be 
$1/6, 2/9, 1/6, 4/9$, respectively.
From this we find that 
\[H(B) = 15/9~log 3 - 7/9 \approx 1.864 ~bits.\]
Likewise, the joint entropy, $H(A,B)$,  is easily computed from
the table of probabilities.
\[H(A,B) = 15/9~log 3 + 5/9 \approx 3.197 ~bits.\]

In other words, the amount of information 
leaking in this system is 
$I(A;B) = log 3 - 4/3 \approx 0.252$ bits/photon, or
about $16\%$.
Or the actual uncertainty from the point of Bob or Eve
is $log 3 - 0.252 = 4/3$ bits per photon.

Since the k-photons (that constitute 1 out of every
3 photons transmitted) do not ultimately contribute to the
formation of the key, the information being sent out by
Alice is at the rate of $2/3 \times 4/3 = 8/9$ bits.
With a $50 \%$ uncertainty for each photon, only half
of them wil lead to the key sequence, so the information
rate is now $1/2 \times 8/9 = 4/9$ bits, which is
precisely the value we have argued using a different
reasoning.
This analysis confirms the information exchange figures
associated with
our protocol.

\section{When n photons are transmitted}

When n photons are transmitted we will, on an average,
be able to obtain $\frac{4n}{9}$ bits for the key
and an additional $\frac{n}{9}$ bits for
authentication of the transmission process.
In other words, the transmission would be certified to the probability
$1-3^{-n/9}$.

In comparison, the 4-state BB84 protocol provides $\frac{n}{2} - m$
key bits and certification of 
$1-2^{-m}$.
When $n/2 - m = 4n/9$, the key bits are the same
number for the two cases, or for $n = 18 m$.

We obtain more key bits from the 3-state protocol compared
to the BB84-protocol for $n < 18 m$.
Thus for $n= 54$ and $m=6$, the 3-state protocol gives
24 usable bits, while the BB84-protocol gives only 21 usable bits.

Furthermore, the certification probability is higher.
The two certification probabilities will be the same if
\[ 1-3^{-n/9} = 1-2^{-m}.\] 
Or BB84-certification probability is
higher for $m > 0.176 n$. But this is an impossible range
of comparison because the effective value of $m$ in the
3-state protocol is  only $ n/9$.

We stress again that the certification probability
comparisons are only notional, based on our assumption of
a 3-state detection scheme used by Eve.
Real comparison would require that the nature of the
attack strategy be spelt out.

\section{Concluding remarks}
By weakening some assumptions regarding the nature of
the quantum cryptographic system, 
we have devised a new  3-state protocol that
has some attractive features, and it
provides unexpectedly good 
results. 
It has the unique
property that no further exchange of verification information
is necessary after the initial steps of the protocol, unless
one desires the certification probability to be greater than
what is inherent in the system.

Quantum key distribution systems described here and
elsewhere (Bennett 1992; Ekert 1991) are fundamentally dual in nature
in as much that part of the information must be sent over
a classical channel, as happens in the post-photon
transmission communications between Alice and Bob.
This aspect makes the communication system somewhat similar
to those brain models that postulate an underlying
quantum basis to cognitive processes (e.g. Hameroff 1998; Kak 1996).
Is the ``understanding'' of the incoming sensory stimuli facilitated by
the filtering information mapped into the neural
structures of the brain during the process of evolution?
And if the universe of the stimuli is itself structured, as is
the case for the polarizations chosen by Alice, 
then is the subject led to an understanding of this
structure by trying out various ``stuck'' settings
of his receiver apparatus and exploiting
the fact that biological signals are often repetitive?

\section*{References }
\begin{description}

\item
Bennett, C H and Brassard, G 1984
``Quantum cryptography:
Public key distribution and coin tossing,'' 
{\em Proceedings of the IEEE Intl.
Conf. on Computers, Systems, and Signal Processing}, Bangalore, India
(IEEE New York, 1984, pages 175-179).

\item
Bennett, C H 1992
{\em Phys. Rev. Lett.} {\bf 68}, 3121 

\item
Ekert, A K 1991
{\em Phys. Rev. Lett.} {\bf 67}, 661

\item
Hameroff, S 1998
{\em Phil. Trans. R. Soc. Lond. A} {\bf 356}, 1869 

\item Kak, S 1985
{\em IEEE Trans. on Computers}  {\bf C-34}, 803 

\item
Kak, S 1996
 ``The three languages of the brain: Quantum, reorganizational,
and associative.'' In
{\it Learning as Self-Organization},
K. Pribram and J. King (eds.).
(Lawrence Erlbaum Associates, Mahwah, 1996, pages 185-219).


\item
Kak, S 1998
{\it Foundations of Physics} {\bf 28}, 1005

\end{description}
 
\end{document}